\documentclass[pre,twocolumn]{revtex4}
\usepackage{epsfig}
\usepackage{bm}
\usepackage[latin1]{inputenc}

\begin{document}
\title{Dimensions, Maximal Growth Sites and Optimization
in the Dielectric Breakdown Model}
\author{Joachim Mathiesen$^1$, Mogens H. Jensen$^2$, and Jan {\O}ystein Haavig Bakke$^3$}
\affiliation{$^1$ Physics of Geological Processes, University of Oslo, Oslo, Norway\\ $^2$ The Niels Bohr Institute, Blegdamsvej, Copenhagen, Denmark\\ $^3$ Department of physics, Norwegian University of Science and Technology, Trondheim, Norway}
\email{joachim.mathiesen@fys.uio.no; mhjensen@nbi.dk}
\begin{abstract}
We study the growth of fractal clusters in the Dielectric
Breakdown Model (DBM) by means of iterated conformal mappings. In
particular we investigate the fractal dimension and the maximal
growth site (measured by the Hoelder exponent $\alpha_{min}$) as a
function of the growth exponent $\eta$ of the DBM model. We do not
find evidence for a phase transition from fractal to non-fractal
growth for a finite $\eta$-value. Simultaneously, we observe that
the limit of non-fractal growth ($D\to 1$) is consistent with $\alpha_{min}
\to 1/2$. Finally, using an optimization principle, we give a recipe on how
to estimate the effective value of $\eta$ from temporal growth
data of fractal aggregates.

\end{abstract}

\maketitle
\section{Introduction}
Laplacian growth and the formation of complex patterns has been
the subject of numerous theoretical and experimental works. The
classical examples are the ramified pattern appearing in a
Hele-Shaw cell when a less viscous fluid is injected into a more
viscous fluid \cite{Feder} and the fractal structures emerging
from the particle aggregation in Diffusion-limited Aggregation
(DLA) \cite{81WS}. In the latter example mono-disperse particles
are released one-by-one from a remote source and diffuse until
they hit and irreversibly adhere to a seed cluster at the center
of coordinates. The cluster slowly expands as particles are added.
Statistically, the motion of a single particle is described by the
harmonic potential $U$ satisfying the Laplace equation $\Delta
U=0$ and the probability of sticking to the cluster at a specific
site, $z$, is given by the harmonic measure $|\nabla U (z)|$. The
formulation of DLA is contained within a more general model, the
Dielectric Breakdown Model (DBM) \cite{84NPW,01Ha}, where the
growth probability $\rho_\eta$ at the cluster interface is
proportional to the harmonic measure raised to a power $\eta$,
$\rho_\eta\propto|\nabla U|^\eta$. Despite intensive research in
Laplacian growth, fundamental questions regarding the scaling
properties still have no answer. The growth laws of DLA and DBM
are extremely simple and in apparent disparity to the complex
patterns they produce. The complex patterns arise from a strong
correlation between the position of already aggregated particles
and the influx of new particles. As the outermost tips advance the
probability for particles to reach the parts left behind
diminishes and the harmonic measure broadens and becomes even
multifractal \cite{02JLMP}. For increasing values of $\eta$ the
growth probability will concentrate around the tips and the
fractal dimension gets closer to unity and ultimately, in the
limit of infinite $\eta$, the particle cluster loses fractality.
Recently, it has been speculated that in two dimensions this
transition from fractal to non-fractal growth may happen at a
finite critical value of $\eta$ and numerically, this value has
been found to be $\eta\approx 4$ \cite{93SGSHL,01Ha}.  In the
vicinity of such a critical point it may be safe to disregard the
noise giving rise to local density fluctuations along the branches
\cite{02H}. For that reason, the dominating stochastic component
in the cluster growth is the rate at which growing tips split in
two or more branches. While growing, neighboring branches compete
and if one branch quickly dies after a tip-splitting the growth
will stay non-fractal. It has been shown \cite{01H} that in the
idealized case of straight growing branches, tip-splitting is
suppressed for $\eta>4$ supporting that $\eta_c=4$. Based on the
idealized branch growth model a renormalization group approach has
been used in an expansion around $\eta_c$. Although an expansion
provides important information for small values of $4-\eta$ it may
provide little information on DLA ($\eta=1$). In this article we
test the hypothesis of a critical point at $\eta=4$ performing
extensive numerical simulations. We provide detailed figures on
the dependence of the fractal dimension, $\alpha_{min}$ and the
exponent $\eta$. Moreover, we propose a method for extracting
effective $\eta$-exponents given either experimental or numerical
data series. For that purpose we make use of iterated conformal
maps \cite{98HL} which have proven a convenient tool for
generating conformal mappings of domains of arbitrary shape
\cite{06MPST}, see section \ref{icm}. In Section \ref{opti} a
method is proposed for extracting effective $\eta$ exponents by
optimization. In section \ref{frac} we present results {\em pro et
con} a phase transition in DBM, the maximal growth sites and the
fractal dimensions.

\section{Iterated conformal mappings}\label{icm}
The conformal invariance of the Laplace equation reduces the problem of finding the harmonic measure, $\rho_1(z)$, around any simply connected domain in the complex plane to that of finding a conformal transformation $\omega = \Phi^{-1}(z)$ of the domain to the unit disc
\begin{equation}
\rho_1(z)=\frac 1 {\left | \Phi^{'}(w)\right|}
\end{equation}
The method of iterated conformal mappings provides a general
framework to construct such transforms as well as a simple
procedure to grow DLA clusters. Assume that a DLA cluster of $n$
particles is mapped to the unit disc by $\Phi_n^{-1}$. An extra
particle is added to the cluster by first adding a small bump of
size $\sqrt \lambda$ to the unit disc using a mapping
$\varphi_{n+1}$ and subsequently applying the inverse mapping
$\Phi_n$. Finally, the composed mapping $\Phi_n\circ
\varphi_{n+1}$ transforms the unit disc into a cluster of $n+1$
particles. The basic mapping $\varphi_{n+1}$ is defined by two
parameters the position and size of the bump, the position,
$e^{i\theta}$, is random in DLA since the measure is uniform
around the circle. The size $\sqrt\lambda$ of the $n$'th bump is
controlled by the condition that
$$
\sqrt{\lambda_0}=\sqrt\lambda_n | \Phi' (e^{i\theta})|
$$
Consequently, the particles (transformed bumps) will all to linear order have the same size $\sqrt{\lambda_0}$. The full recursive dynamics is written as iterations of the basic map
\begin{equation}
  \label{eq:4}
  \Phi^{(n)}(w)=\varphi_{\theta_{1},\lambda_{1}}\circ\ldots\circ\varphi_{\theta_n,\lambda_n}(w) \ .
\end{equation}
Note that this structure is unusual in the sense that the order of iterates is inverted compared to standard dynamical systems.

For DBM the growth measure along the cluster interface, parameterized by $s$, is given by
\begin{equation}
\rho_\eta(s)=\frac{\rho_1^\eta(s)}{\int\rho_1^\eta(t) dt},
\end{equation}
 which for $\eta\neq1$ is not conformally invariant. On the unit circle, parameterized by $\theta$, the growth measure transforms into
\begin{equation}
  \label{eq:5}
  \rho_\eta(\theta)d\theta \sim
  \rho_\eta(s(\theta))\left|\frac{ds}{d\theta}\right|d\theta\sim
  |\Phi'(e^{i\theta})|^{1-\eta}d\theta
  \ .
\end{equation}
In the simulations we choose $\theta$ according to the distribution $\rho_\eta$ using standard Monte Carlo samplings of the measure. The number of samples needed for an accurate estimate of the distribution increases with $\eta$ and is chosen according to
\begin{equation}
  \label{mc}
\frac k {\sqrt{ \lambda_0}\max_s \rho_1(s)}
\end{equation}
By choosing $k>1$, the site of maximal measure will on the average be visited more than once during the sampling. It turns out that there is no visible change in the scaling of the clusters when choosing $k>1$, see Fig. \ref{conv} for a test of convergence as function of $k$; in the results presented below, we use $2\leq k\leq 8$.

\begin{figure}[htbp]
  \epsfig{file=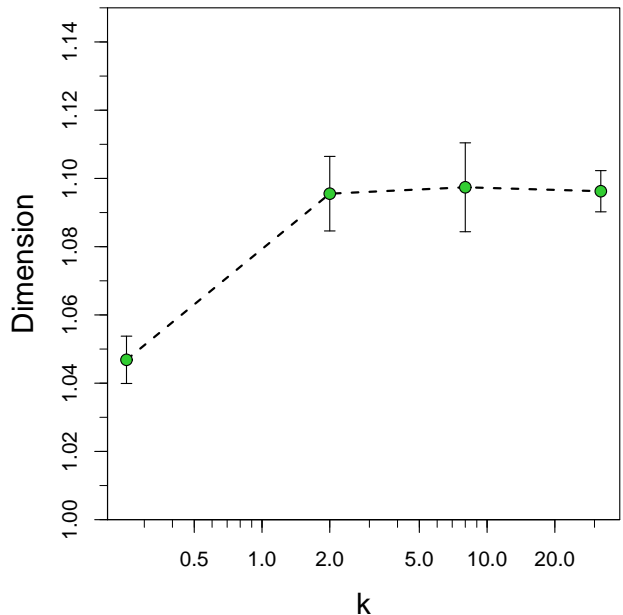,width=.475\textwidth}
  \caption{(Color online) Test of convergence for the deployed Monte Carlo method. Estimates of the fractal dimension for $\eta=4$ and for Monte Carlo samples given by $k=1/4, 2, 8, 32$ in Eq. $(\ref{eq:5})$. For each data point, we used four clusters of size $40000$ particles.}
  \label{conv}
\end{figure}

\section{Extracting effective $\eta$ exponents by optimization}\label{opti}
Consider an interface growing at a rate determined by some unknown function of the harmonic measure. The method of iterated conformal mappings is readily turned into a framework for estimating this function. More specifically, it is here demonstrated on numerical simulation data of the DBM that the value of $\eta$ can be extracted from a careful tracking of the cluster growth. The general idea is to utilize the iteration scheme in tracking the motion of the interface by gradually expanding the mapping, see \cite{06MPST} for further details. The harmonic measure is recorded as the interface evolves and from a maximum likelihood principle the $\eta$ value of the growth is extracted.
The probability for growth to occur at a site $z_n$ at the interface is in a given growth step $n$ approximated by the sum
\begin{equation}
  \label{etaprob}
  \rho_\eta(n,z_n)=\frac{\rho_1^\eta(n,z_n)}{\sum_z \rho_1^\eta(n,z)}
\end{equation}
From this expression, more ways exist to estimate the $\eta$ value used in the simulation. Assuming that the $n$'th growth event occurred at the site $z_n$, a direct estimate of $\eta$ follows from maximizing $\rho_\eta(z_n)$ with respect to $\eta$. Naturally, this will lead to dramatic fluctuations in the estimates and therefore maximizing products of $\rho_\eta$ over several growth steps provides a better estimate,
\begin{equation}
  \label{est}
  \prod_k \rho_\eta(k,z_k)
\end{equation}

\begin{figure}
\centering
\epsfig{width=.45\textwidth,file=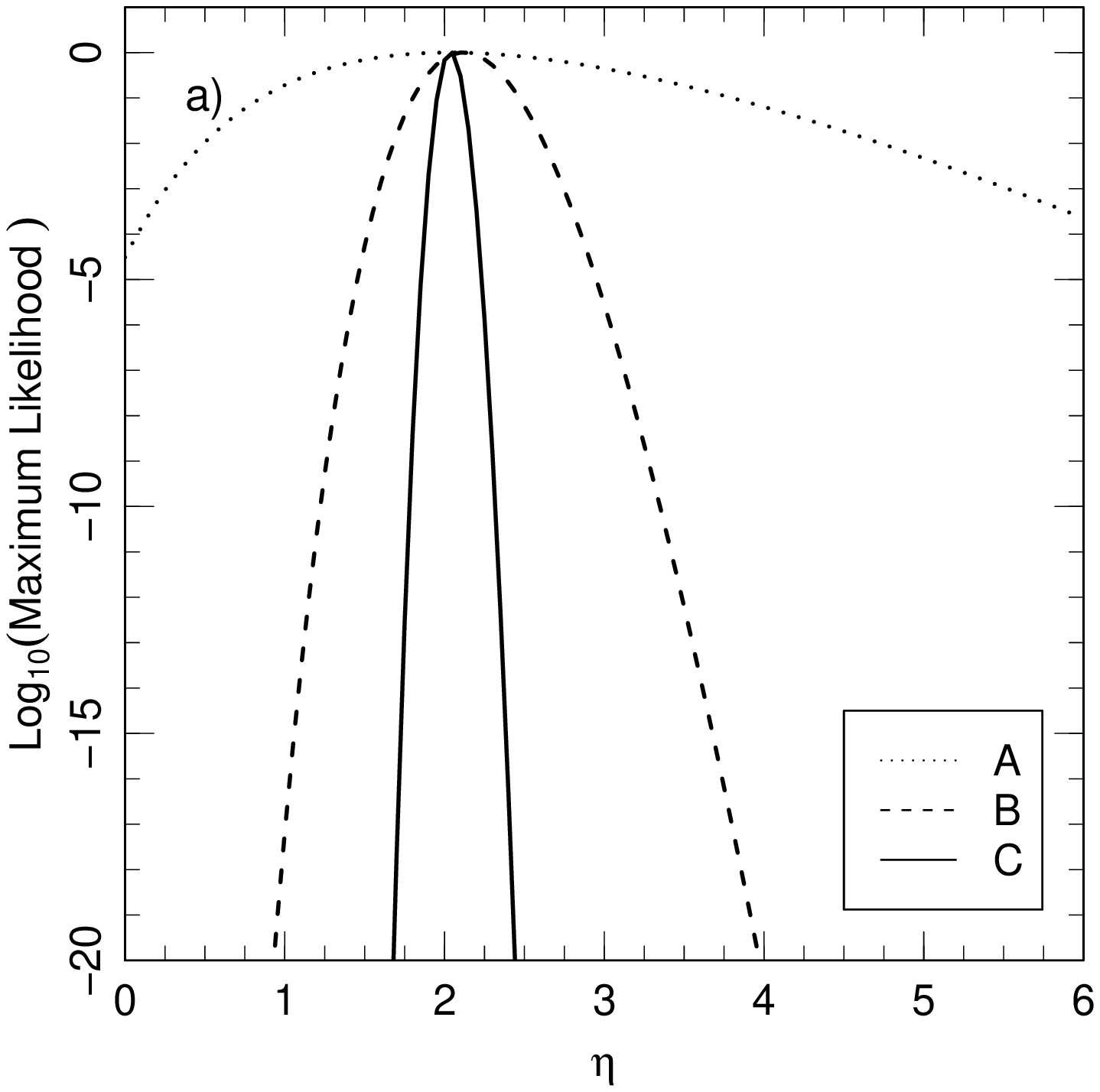}
\epsfig{width=.45\textwidth,file=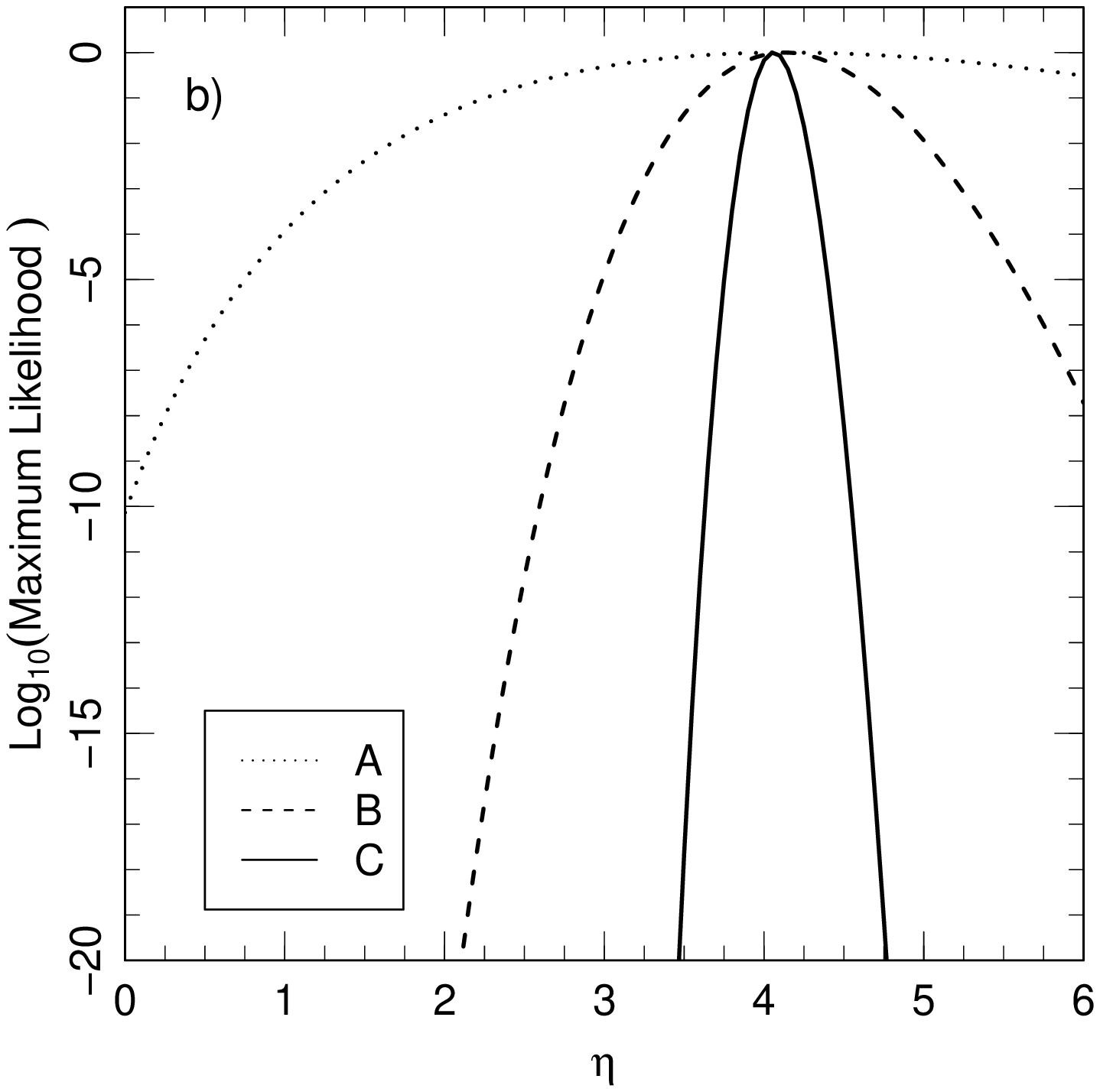}
\caption{Product of the growth measure, maximum likelihood, $\prod_{k=1}^{N}
  \rho_\eta(2k,z_{2k})$ as function of $\eta$, for $N=15,150,1500$ (A,B, and C, respectively). The clusters applied where grown with a) $\eta=2$ and b) $\eta=4$, consistent with the extreme values of the maximum likelihood. Note that the products were normalized by their maximum value and that every second growth step was used only. The latter was done to avoid the products becoming too small.
}\label{optimum}
\end{figure}

In Fig. \ref{optimum}, we show how this product varies as function of $\eta$ and with the number of factors used. With an increasing number of factors the maximum becomes more pronounced and the $\eta$ value used in the simulations is easily recovered. These products confirms that the number of Monte Carlo samples used in Eq. (\ref{mc}) are appropriate and more importantly that the method is directly applicable to experimental data for estimating an effective $\eta$ value or more generally the boundary condition function determining the growth rate.


\section{Dimension and $\alpha_{min}$}\label{frac}
\begin{figure}[htbp]
  \epsfig{file=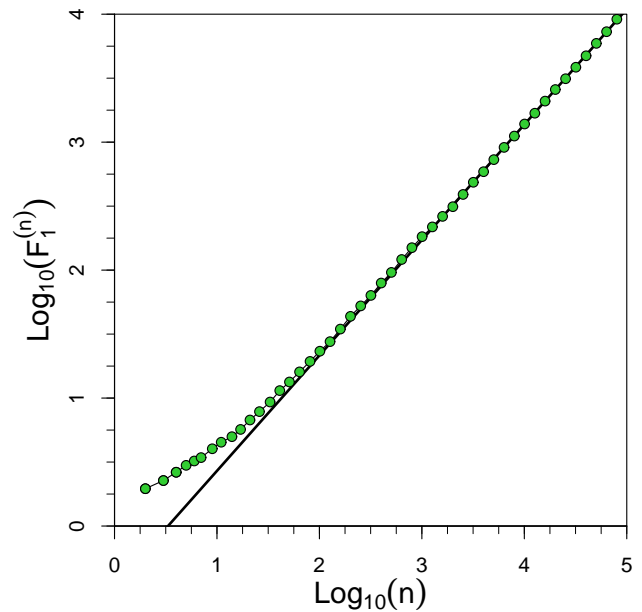,width=.475\textwidth}
  \caption{(Color online) First Laurant coefficient versus cluster size for $\eta=4$ and 80000 particles (with $k=2$, see text). The added line is a fit of the fractal dimension $D=1.10$}
  \label{laurant}
\end{figure}

\begin{figure}[htbp]
  \epsfig{file=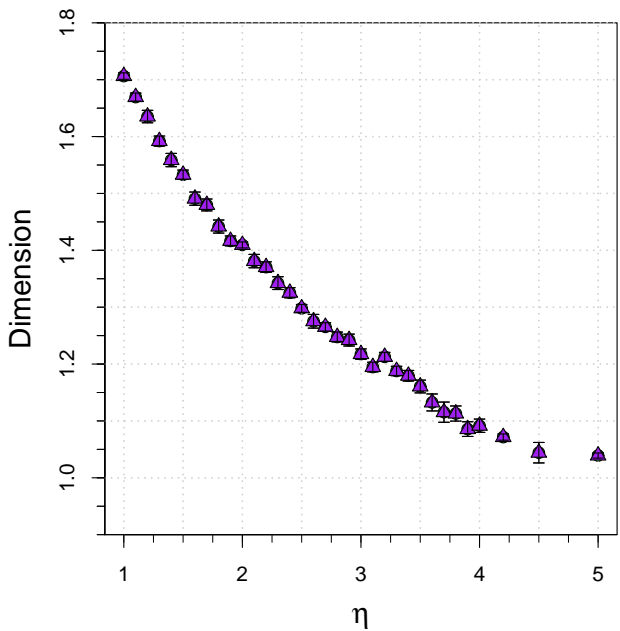,width=.475\textwidth}
  \caption{(Color online) Computed values of the fractal dimension versus $\eta$. The dimension was extracted from a fit of the first Laurant coefficient $F_1$ using clusters of sizes 20000-40000 particles for $\eta\leq 3.5$ and 40000-80000 particles for $\eta> 3.5$. Each data point is averaged over 20 clusters and the error bars are estimated by the standard deviation.}
  \label{dim}
\end{figure}

The dimension of a cluster grown by this conformal mapping
technique is determined by the first term in the Laurant expansion
of $\Phi^{(n)}$, $F_1^{(n)}$, which will scale like $F_1^{(n)}
\sim n^{1/D} \sqrt{\lambda_0}$ \cite{98HL}. The dimension is thus
estimated by a direct fit of this scaling law as demonstrated in Fig. \ref{laurant} for a cluster 80000 particles and $\eta=4.0$. 
Using the conformal
mapping technique we have grown clusters up to sizes 80000
particles with varying values of $\eta$ in the interval $\eta \in
[1,5]$. Fig. \ref{dim} shows the results for the value of the
dimension versus $\eta$. As is clear from the figure, the
value of the dimension decreases smoothly with $\eta$, from the
DLA value $D=1.71$ for $\eta=1$ down towards $D \sim 1$ for $\eta
\to \infty$. Hastings \cite{01H} presented arguments in favor of
an upper critical dimension $\eta_c = 4$ for which the clusters
become one-dimensional. We however do not observe indications of
this transition. As seen in Fig. \ref{dim} it is quite clear that
the data smoothly bends away before reaching the point
$(\eta,D)=(4,1)$ and only approaching the one-dimensional growth
in the limit of large $\eta$-values. We thus conclude that there
do not exist a critical point at a finite $\eta$.

\begin{figure}
\centering
\epsfig{width=.45\textwidth,file=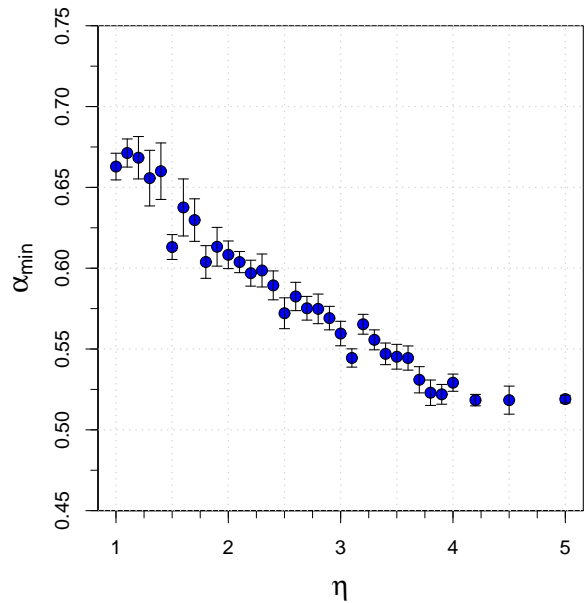}
\caption{(Color online) Computed values of $\alpha_{min}$ versus $\eta$ using the same clusters as in Fig. \ref{dim}. 
}\label{amin}
\end{figure}

Halsey \cite{02H} has computed a first-order correction to $D$ for
$\eta<4$, obtaining $D= 1 + \frac{1}{2} (4-\eta) + O(4-\eta)^2$.
This relation predicts a linear variation of slope $\frac{1}{2}$
around $\eta_c = 4$. As seen in Fig. \ref{dim} we do not observe this
behavior. 

It is well know that the growth measure of a DBM model exhibits
multifractal properties with a spectrum of growth exponents
measured by local Hoelder exponents $\alpha$ \cite{02JLMP}. The points
of highest growth measures are characterized by the minimum $\alpha$-value,
$\alpha_{min}$. We have earlier determined this value using the
iterated conformal mapping technique \cite{03JMP} and extend it here
to the DBM model. In this method, it is very easy to keep
track of where the maximum growth probability is located as more particles are added.
Let us assume that at the ($n$-1)'th growth step the site
with the largest probability is located at the angle
$\theta_{max}$ on the unit circle, i.e. for all
$\theta$
\begin{equation}
\frac 1 {|{\Phi^{(n-1)}}' (e^{i \theta_{max}})|}\geq
\frac 1 {|{\Phi^{(n-1)}}' (e^{i \theta})|}
\end{equation}
When we add a new bump in the $n$'th growth step the position
of maximal probability may not change (up to reparameterization
of the angle $\theta_{max}$), or move to the new bump.
We can easily find the reparameterized angle and determine
the new position from
\begin{equation}
\label{pmaxn}
  \rho_1^{max,n}=\max\left\{\frac 1 {|{\Phi^{(n)}}'
(\phi^{-1}_{\lambda_n,\theta_n}(e^{i\theta_{max}}))|},
\frac 1 {|{\Phi^{(n)}}' (e^{i \theta_n})|}\right\}\ .
\end{equation}
If $\rho_1^{max,n}$ is located at $\theta_n$ we put
$\theta_{max}=\theta_n$ in the $(n+1)$'th growth step.
Using conformal mappings, we have also previously estimated the
critical branching angle as a function of $\eta$ in the DBM model \cite{02MJ}.
\begin{figure}
\centering
\epsfig{width=.45\textwidth,file=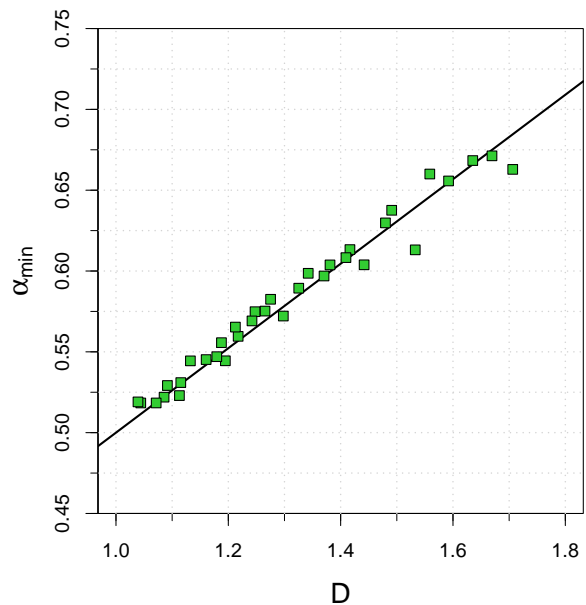}
\caption{(Color online) $\alpha_{min}$ versus the fractal dimension $D$. The data are the same as used in Figs. \ref{dim} and \ref{amin}
}\label{dimamin}
\end{figure}

Fig. \ref{amin} shows the results of $\alpha_{min}$ vs. $\eta$ and
we observe that $\alpha_{min}$ decreases from the DLA values
$\alpha_{min} =0.68$ down to $\alpha_{min} =0.5$. It is obvious
that $\alpha_{min} =0.5$ corresponds to the Hoelder exponent for a
line. In consistency with the results in Fig. \ref{dim} we observe that the
curve bends smoothly and that the one-dimensional growth is only
obtained in the limit $\eta \to \infty$. The last figure, Fig. \ref{dimamin}, shows $\alpha_{min}$ plotted
vs. $D$. By extrapolation (as indicated by the line) we see that
$\alpha_{min}$ assumes its minimal value 0.5 at a dimension $D
=1.0$.

\section{Conclusions}
The conclusions of our paper are twofold. Firstly, we have
presented a method to extract the effective value of the growth
exponent $\eta$, for a time series of a growing aggregates,
assuming an underlying mechanism based on the Dielectric Breakdown
Model (DBM) model. The estimate is based on a maximum likelihood
method and converges rather well for the numerical data presented
here. We believe this method should be directly applicable to
experimental data when it is possible to extract intermediate
steps in the formation of the aggregates. We urge the method to be
used in for example viscous fingering experiments in random media
\cite{04L}. Secondly, we have thoroughly investigated the scaling
structure of DBM clusters as a function of the growth exponent
$\eta$. Based on extensive numerical simulations we do not find
support for the conjecture that the growth becomes
one-dimensional at the critical value $\eta_c = 4$\cite{01H,02H}. On the contrary, our results
indicate that there do not exist a critical point for at finite
$\eta$-value and that the scaling exponent
of the maximal growth site $\alpha_{min}$ assumes its
minimal value 0.5 when the growth becomes non-fractal.

\section{Acknowledgements}
We thank Knut Joergen Maaloy and Stephane Santucci for interesting
discussions at an early stage of this work. This project was funded by \textsl{Physics of Geological Processes},
a Center of Excellence at the University of Oslo, the Danish National Research Foundation and the
VILLUM KANN RASMUSSEN Foundation for support.



\end{document}